\documentclass{PoS}
\usepackage{amsmath}
\usepackage{amssymb}
\usepackage{graphicx}
\usepackage{cite}
\usepackage{tabularx}
\PoS{PoS(LAT2005)289}

\title{Deconfinement phase transitions in external fields}

\ShortTitle{Deconfinement in external fields}

\author{\speaker{Paolo Cea} \\ Dipartimento
Interateneo di Fisica, Universit\`a di Bari  and INFN - Sezione di Bari, \\
I-70126 Bari, Italy \\
E-mail:  \email{Paolo.Cea@ba.infn.it} }

\author{Leonardo Cosmai \\ INFN - Sezione di Bari, I-70126 Bari,
Italy\\
E-mail:  \email{Leonardo.Cosmai@ba.infn.it}}

\abstract{
We compare vacuum dynamics of abelian versus non abelian lattice gauge theories
in presence of external fields. We find that the deconfinement temperature
of non abelian  theories depends on the strength of a constant abelian chromomagnetic
field end eventually goes to zero. On the contrary such an effect is not shared
by abelian theories. We argue on the relevance of this result
to understand the QCD vacuum.
}

\FullConference{XXIIIrd International Symposium on Lattice Field Theory\\
         25-30 July 2005\\
         Trinity College, Dublin, Ireland}

\begin{document}

\newcommand{\be}{\begin{equation}}
\newcommand{\ee}{\end{equation}}

\section{Introduction}
\label{Introduction}
A fundamental problem in high energy physics is to understand color confinement. Indeed, the mechanism that leads to color
confinement remains an open question despite intense lattice studies for nearly three decades (for recent reviews on
confinement see~\cite{Ripka:2003vv,Greensite:2003bk,Haymaker:1998cw}).
In order to investigate vacuum structure of  lattice gauge theories both at zero and finite temperature, we introduced a
lattice gauge invariant effective action $\Gamma[\vec{A}^{\text{ext}}]$  for gauge systems in external static background
fields $\vec{A}^{\text{ext}}$~\cite{Cea:1999gn}. If we now consider the gauge theory at finite temperature $T=1/(a L_t)$
in presence of an external background field, the relevant quantity turns out to be the free energy functional
${\mathcal{F}}[\vec{A}^{\text{ext}}]$. We are interested in vacuum dynamics of  U(1), SU(2), and SU(3) lattice gauge
theories under the influence of an abelian chromomagnetic background field. In our previous studies we found that in SU(3)
the deconfinement temperature depends on the strength of an applied external constant abelian chromomagnetic field. This
is at variance of abelian magnetic monopoles where the abelian monopole background fields do not modify the deconfinement
temperature. We would like to ascertain if the dependence of the deconfinement temperature on the strength of an applied
external constant abelian chromomagnetic field is a peculiar feature of non abelian gauge theories. Since our lattice has
the topology of a torus, the magnetic field turns out to be quantized $ a^2 \frac{g H}{2} = \frac{2 \pi}{L_1}
n_{\text{ext}}$, $n_{\text{ext}}\,\,\,{\text{integer}}$. Moreover, since the free energy functional
${\mathcal{F}}[\vec{A}^{\text{ext}}]$ is
invariant for time independent gauge transformations of the background field $\vec{A}^{\text{ext}}$, it follows that for a
constant background field  the free energy $F[\vec{A}^{\text{ext}}]$ is proportional to the spatial volume $V=L_s^3$, and the relevant
quantity is the density of free energy $ f[\vec{A}^{\text{ext}}] = \frac{1}{V} F[\vec{A}^{\text{ext}}]$. We evaluate by
numerical simulations $f^{\prime}[\vec{A}^{\text{ext}}]$ the derivative with respect to the coupling $\beta$ of the free
energy density $f[\vec{A}^{\text{ext}}]$ at fixed external field strength $gH$.
%
\section{(3+1) dimensions}
\label{4dim}
In this section we report results obtained (see also Ref.~\cite{Cea:2005td})
in studying the finite temperature phase transition of lattice gauge theories
SU(3) and  SU(2) in (3+1)-dimensions, in presence of a constant abelian chromomagnetic background field.
%
%
\subsection{SU(3)}
\label{4dimSU3}
As is well known, the pure SU(3) gauge system undergoes a deconfinement phase transition at a given critical temperature.
Our aim is to study the possible dependence of the critical temperature from the strength of the applied field. The
critical coupling $\beta_c$ can be evaluated by measuring $f^{\prime}[\vec{A}^{\text{ext}}]$, the derivative of the free
energy density with respect to $\beta$, as a function of $\beta$. Indeed we found that $f^{\prime}[\vec{A}^{\text{ext}}]$
 displays a peak in the critical region where it can be parameterized as
 $\frac{f^{\prime}(\beta,L_t)}{\varepsilon^{\prime}_{\text{ext}}} = \frac{a_1(L_t)}{a_2(L_t) [\beta -
\beta^*(L_t)]^2 +1} $, where we normalize   to the derivative of the classical energy due to  the external applied field
$\varepsilon^{\prime}_{\text{ext}} = \frac{2}{3} \, [1 - \cos( \frac{g H}{2} )] = \frac{2}{3} \, [1 - \cos( \frac{2
\pi}{L_1} n_{\text{ext}})]$. Remarkably, we have checked that the evaluation of the critical coupling $\beta^*(L_t)$ by
means of $f^{\prime}[\vec{A}^{\text{ext}}]$ is consistent with the usual determination obtained through the temporal
Polyakov loop susceptibility. We varied the strength of the applied external abelian chromomagnetic background field to
study quantitatively the dependence of $ T_c$ on $gH$. It is worth to note that lattice data can be reproduced by the
linear fit  $\frac{T_c}{\sqrt{\sigma}} =  \alpha \frac{\sqrt{gH}}{\sqrt{\sigma}} + \frac{T_c(0)}{\sqrt{\sigma}} $ with $
\frac{T_c(0)}{\sqrt{\sigma}} = 0.643(15)$, $ \alpha = -0.245(9)$.  The critical field can now be expressed in units of the
string tension $ \frac{\sqrt{gH_c}}{\sqrt{\sigma}} = 2.63  \pm 0.15$. Assuming $\sqrt{\sigma}=420$~MeV, we find for the
critical field $\sqrt{gH_c} = (1.104 \pm 0.063) {\text{GeV}}$ corresponding to $gH_c=6.26(2) \times 10^{19}$~Gauss.
%
%
%
%
\FIGURE[ht]{\label{Fig3}
\includegraphics[width=0.58\textwidth,clip]{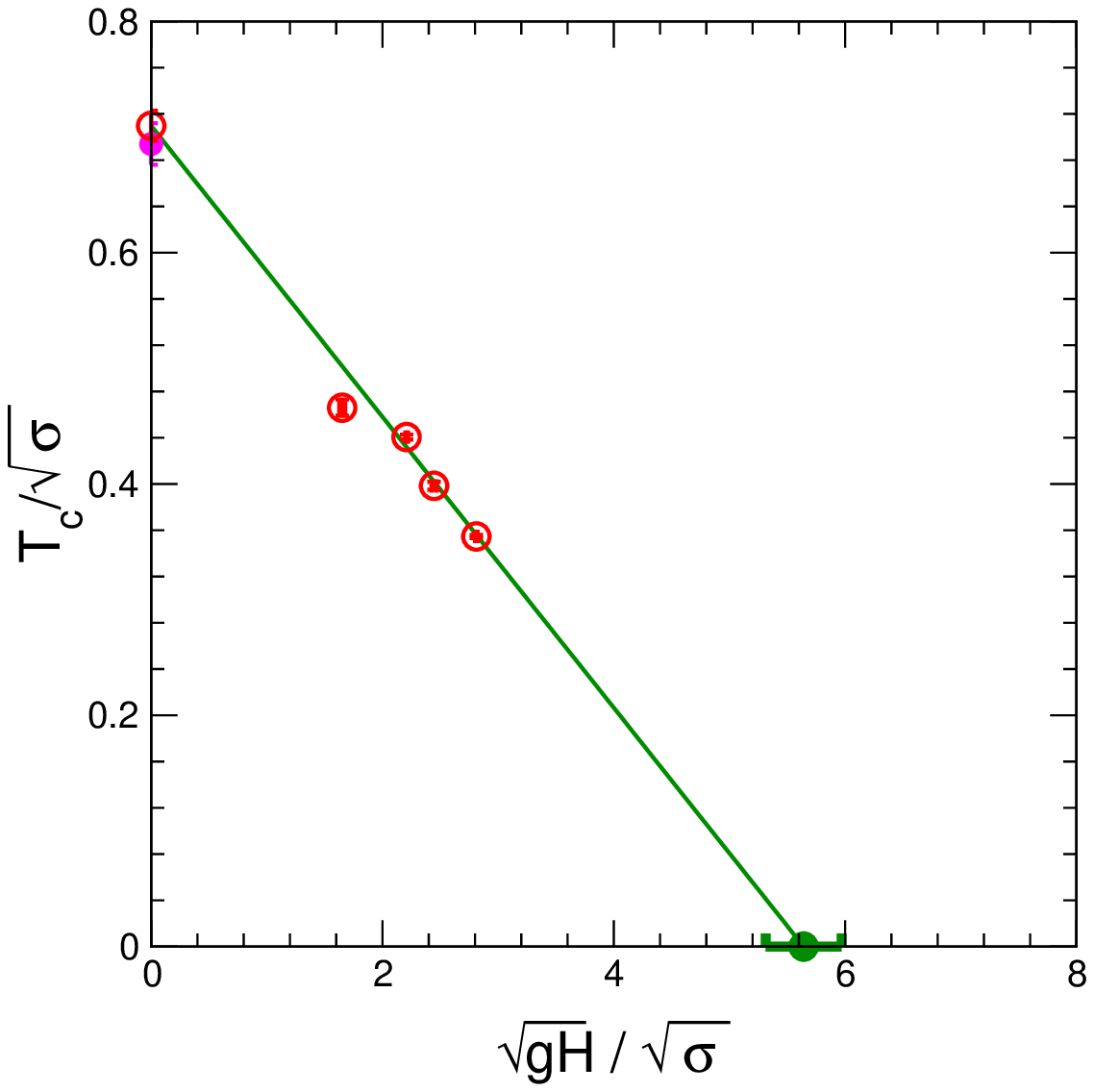}
\caption{SU(2) in (3+1) dimensions. The critical temperature $T_c$ estimated on a $64^3 \times 8$ lattice  versus the
square root of the field strength $\sqrt{gH}$ in units of the string tension. } }
%
%
\subsection{SU(2)}
\label{4dimSU2}
%
We also studied the SU(2) lattice gauge theory in a constant abelian
chromomagnetic field. Even in this theory the deconfinement
temperature turns out to depend on the strength of the applied
chromomagnetic field. We evaluated the critical coupling
$\beta^*(L_t,n_{\text{ext}})$ on a  $64^3 \times 8$ lattice versus
the strength of the external chromomagnetic field. As in previous
section the critical coupling has been found by locating the peak of
the derivative of the free energy density with respect to the gauge
coupling $\beta$. In Fig.~\ref{Fig3}  $T_c/\sqrt{\sigma}$ is plotted
against $\sqrt{gH}/\sqrt{\sigma}$.  As in the SU(3) case discussed
in previous section, we found that the linear fit works quite well
and we get $ \frac{T_c(0)}{\sqrt{\sigma}} = 0.710 (13)$,   $\alpha =
-0.126 (5)$. The value obtained for $T_c(0)/\sqrt{\sigma}$ is in
good agreement with the value $T_c/\sqrt{\sigma}=0.694(18)$, without
external field, obtained in the literature.
%
\FIGURE[ht]{\label{Fig4}
\includegraphics[width=0.60\textwidth,clip]{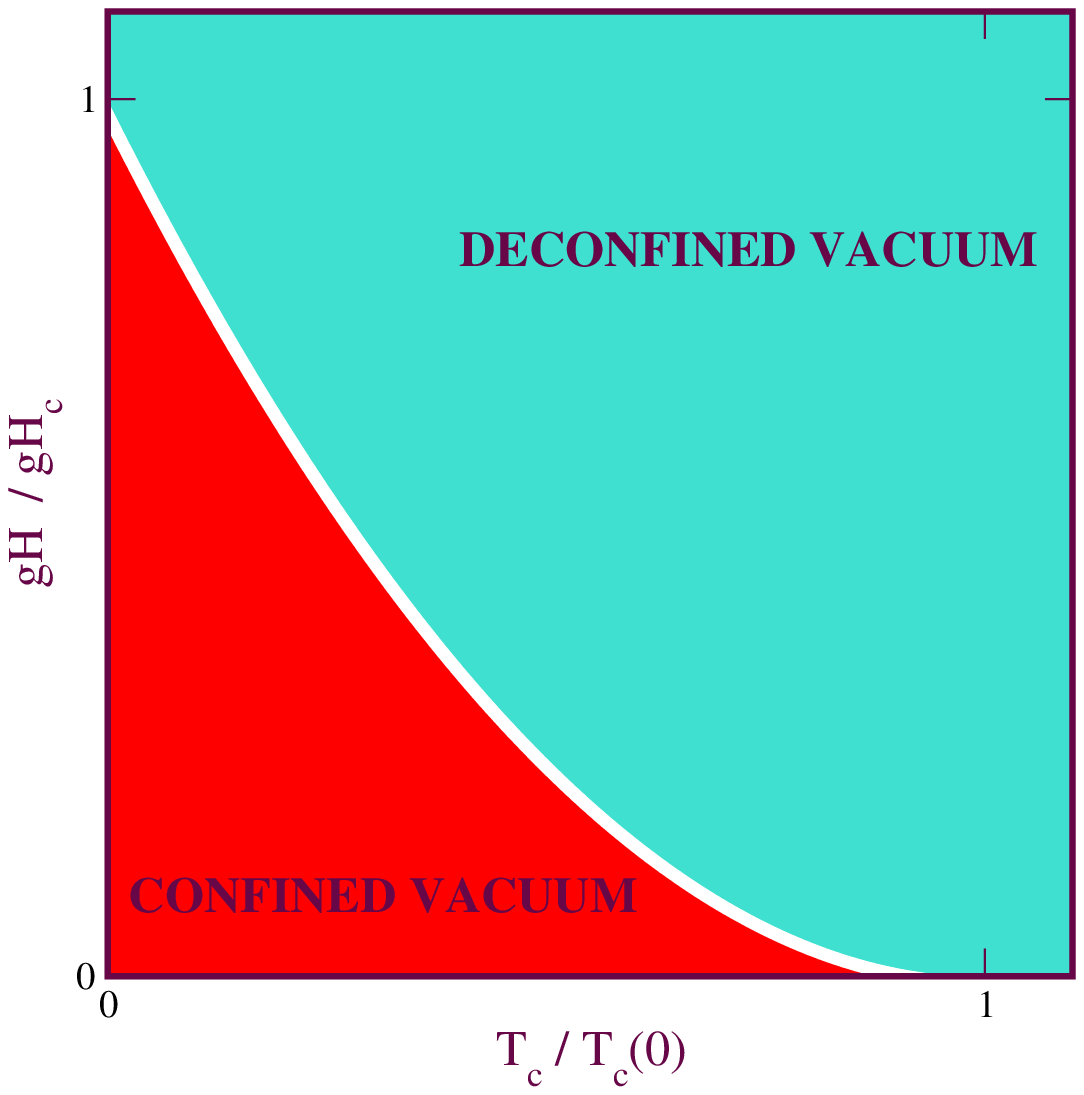}
\caption{Phase diagram of four dimensional $SU(2)$ and $SU(3)$ gauge theories.}}
Now we can estimate the critical field in string tension units that turns out to be $ \frac{\sqrt{gH_c}}{\sqrt{\sigma}} =
5.33  \pm 0.33 $. Note that the critical field $\sqrt{gH_c}/\sqrt{\sigma}$ is about a factor 2 greater than the SU(3)
critical value. Our results indicate a dependence of the deconfinement temperature on the strength of a constant abelian
chromomagnetic background field. On the other hand, we found that such an effect is absent for four dimensional U(1)
lattice gauge theory, so that we may conclude that it  is peculiar of non abelian gauge theories. In Fig.~\ref{Fig4} we
display the phase diagram for four dimensional SU(2) and SU(3) gauge theories.
%
%
\section{(2+1) dimensions}
\label{3dim}
%
%
%
Our numerical results for non abelian gauge theories SU(2) and SU(3) in (3+1) dimensions
in presence of an abelian constant chromomagnetic background field lead us to conclude that the
deconfinement temperature depends on the strength of the applied field, and eventually
becomes zero for a critical value of the field strength.
A natural question arises if this phenomenon, which is peculiar of non abelian gauge theories, continues
to hold in (2+1) dimensions. To this purpose we consider here the non abelian SU(3) lattice gauge theory
to be contrasted with the abelian U(1) lattice gauge theory at finite temperature.
%
\subsection{SU(3)}
\label{3dimSU3}
%
\FIGURE[ht]{\label{Fig6}
\includegraphics[width=0.68\textwidth,clip]{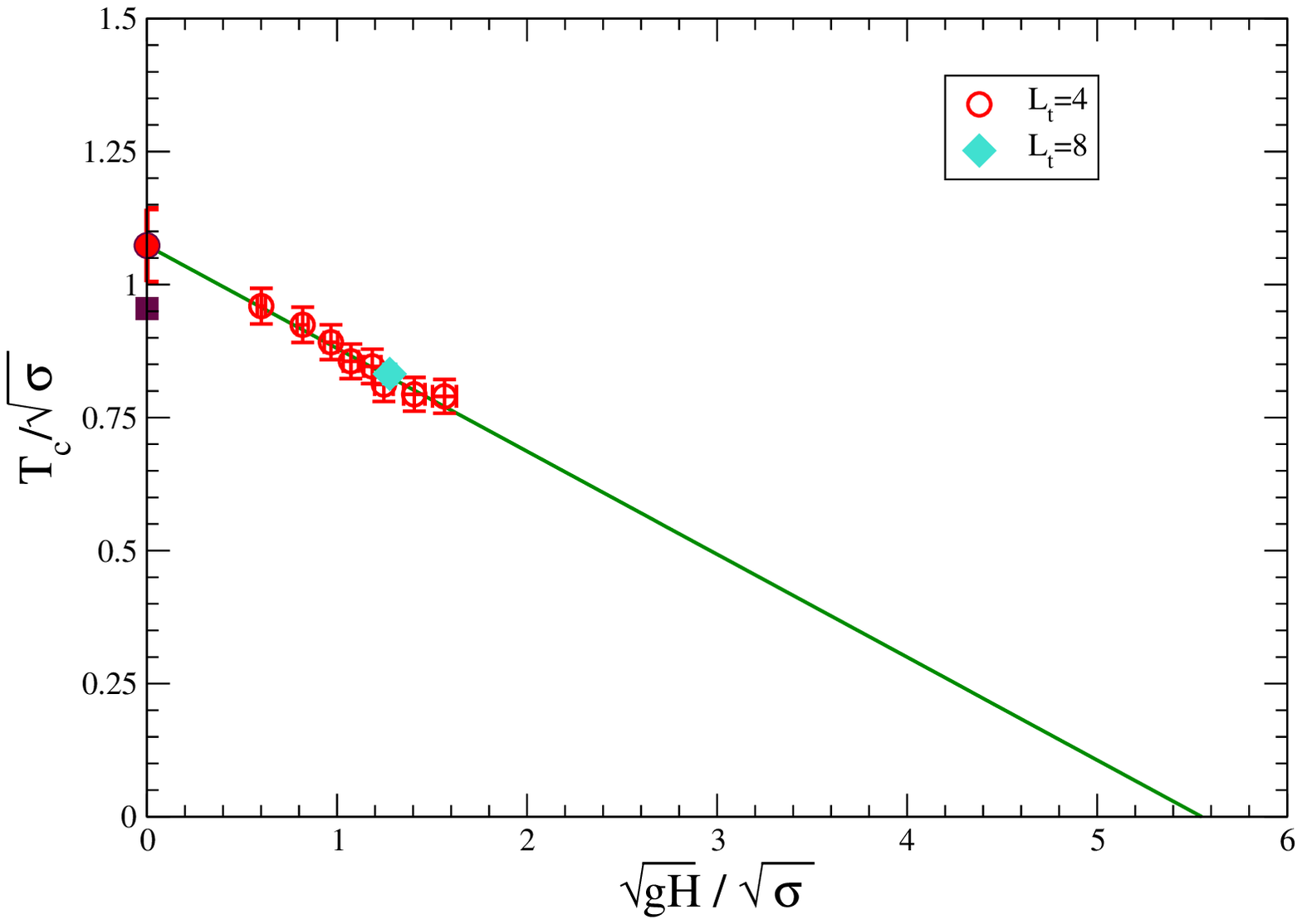}
\caption{SU(3) in (2+1) dimensions. The critical temperature $T_c$ estimated on  $256^2 \times 4$, $512^2 \times 4$ and
$512 \times 256  \times 8$ lattices  versus the square root of the field strength $\sqrt{gH}$ in units of the string
tension. Open circles refer to $L_t=4$, diamond to $L_t=8$. }}
In this section we focus on gauge systems in (2+1) dimensions. As is well known gauge theories in (2+1) dimensions possess
a dimensionful coupling constant, namely $g^2$ has dimension of mass and so provides a physical scale. In (2+1) dimensions
the chromomagnetic field $H^a$ is a (pseudo)scalar $ H^a = \frac{1}{2} \varepsilon_{ij} F^a_{ij} = F^a_{12} $.
As in the four dimensional case since we assume to have a lattice with toroidal geometry the field strength is quantized.
We computed the derivative of the free energy density  on a $L \times 256 \times 4$ lattice, with $L=256, 512$ and several
values of the external field strength parameterized by $n_{\text{ext}}$. We locate the critical coupling $\beta_c$ as the
position of the maximum of the derivative of the free energy density at given external field strength. As for SU(3) in
(3+1) dimensions, the value of $\beta_c$ depends on the field strength. Using the parameterization for the string tension
given in eq.~(C9) of ref.~\cite{Teper:1998te} we are able to estimate the critical temperature $T_c$ in units of the
string tension. We find that, as in (3+1) dimensions, $T_c/\sqrt{\sigma}$ depends linearly on the applied field strength
(see Fig.~\ref{Fig6}). The linear fit  gives $\frac{T_c(0)}{\sqrt{\sigma}}= 1.073 (87)$,  $\alpha=-0.193(76)$ that implies
a critical field $\sqrt{gH_c}/\sqrt{\sigma} = 5.5 \pm 3.7$. Note that the value for $T_c(0)/\sqrt{\sigma}$ in the present
work is in fair agreement with $T_c/\sqrt{\sigma}=0.972(10)$ without external field obtained in ref.~\cite{Engels:1997dz}.
To check possible finite volume effects, we performed a lattice simulation with $L_t=8$. The result shows that within
statistical uncertainties our estimate of the critical temperature from the simulation with $L_t=8$ is in agreement with
result at $L_t=4$.
%
\subsection{U(1)}
\label{3dimU1}
In a classical paper~\cite{Polyakov:1976fu} Polyakov showed that compact quantum electrodynamics in (2+1) dimensions at
zero temperature confines external charges for all values of the coupling. Moreover it is well ascertained that the
confining mechanism is the condensation of magnetic monopoles which gives rise to a linear confining potential and a
non-zero string tension. At finite temperature  the gauge system undergoes a deconfinement transition which appears to be
of the Kosterlitz-Thouless type. We studied lattice U(1) gauge theory in an uniform external magnetic field. We
performed numerical simulations on $512 \times 256 \times 4$ and $512 \times 64 \times 8$ lattices. To determine the
critical coupling $\beta_c$, we measure the derivative of the free energy density. Contrary to the case of (2+1) and (3+1)
non abelian lattice gauge theories, we do not find a dependence of the critical value of the coupling $\beta_c$ on the
magnetic field strength. By increasing the temporal size to $L_t=8$ the critical coupling increases and is still
independent of the external magnetic field strength. Therefore we can conclude that even in (2+1) dimensional case the
critical coupling does not depend on the strength of the external magnetic field as for U(1) lattice gauge theories in
(3+1) dimensions.
%
\section{Conclusions}
\label{Conclusions}
Let us conclude this paper by recalling our main results.
We probed the vacuum of  U(1), SU(2), and SU(3)  l.g.t.'s  by means of an external
constant abelian  (chromo)magnetic field.
We found that, both in (2+1) and (3+1) dimensions:
for non abelian gauge theories  there is a critical field   $gH_c$            such that
for        $gH > gH_c$                   the  gauge systems are in the deconfined phase
(we named this effect "reversible  vacuum color  Meissner effect").
On the other hand we also found that  for abelian gauge theories  the critical coupling
does not depend on the strength  of the external constant magnetic field.

The dependence of the deconfinement temperature on the strength of the abelian chromomagnetic field
in non abelian gauge theories could be intuitively understood by considering that
strong enough chromomagnetic fields  would force
long range color correlations such that the  gauge system gets deconfined.

On a more speculative side, one may thus imagine
the confining vacuum of non abelian gauge theories as a
disordered chromomagnetic condensate which confines color charges
due to both
the presence of a mass gap
and the absence of long range color correlations,
as argued by R. P. Feynman for QCD in (2+1) dimensions~\cite{Feynman:1981ss}.


\providecommand{\href}[2]{#2}\begingroup\raggedright\endgroup

\end{document}